\documentclass[iop]{emulateapj}

\usepackage{xfrac}

\usepackage{graphicx,epsfig,amssymb,layout,verbatim,rotating,calc,mathrsfs,epstopdf}
\usepackage[sumlimits,intlimits,namelimits]{amsmath}
\usepackage{graphics}
\usepackage{rotate}
\usepackage{enumerate}
\PassOptionsToPackage{linktocpage}{hyperref}
\usepackage{natbib}
\usepackage{footmisc}

\def\lsim{\lower 2pt \hbox{$\, \buildrel {\scriptstyle <}\over
         {\scriptstyle \sim}\,$}}
\newcommand\gsim{\buildrel > \over \sim}

\begin{document}

\title{Testing Dissipative Magnetosphere Model Light Curves and Spectra with \textit{Fermi} Pulsars}
\author{Gabriele Brambilla}
\affil{Dipartimento di Fisica, Universit\`{a} degli Studi di Milano,
Via Celoria 16, 20133 Milano, Italy} \affil{Astrophysics Science
Division, NASA/Goddard Space Flight Center, Greenbelt, MD 20771,
USA} \email{gabriele.brambilla@nasa.gov,
gabriele.brambilla1@studenti.unimi.it}

\author{Constantinos Kalapotharakos}
\affil{University of Maryland, College Park (UMDCP/CRESST), College
Park, MD 20742, USA} \affil{Astrophysics Science Division,
NASA/Goddard Space Flight Center, Greenbelt, MD 20771, USA}

\author{Alice K. Harding}
\affil{Astrophysics Science Division, NASA/Goddard Space Flight
Center, Greenbelt, MD 20771, USA}

\and

\author{Demosthenes Kazanas}
\affil{Astrophysics Science Division, NASA/Goddard Space Flight
Center, Greenbelt, MD 20771, USA}

%
%

\begin{abstract}
We explore the emission properties of a dissipative pulsar
magnetosphere model introduced by \cite{PULSARGAMMA}, comparing its
high energy light curves and spectra, due to
curvature radiation, with data collected by the \textit{Fermi LAT}.
The magnetosphere structure is assumed to be near the 
force-free solution. The accelerating electric field, inside the
light-cylinder, is assumed to be negligible, while outside the
light-cylinder it rescales with a finite conductivity ($\sigma$).
In our approach we calculate the corresponding high energy
emission by integrating the trajectories of test particles that
originate from the stellar surface, taking into account both the
accelerating electric field components and the radiation reaction forces.
First we explore the parameter space assuming different
value sets for the stellar magnetic field, stellar period, and
conductivity. We show that the general properties of the model are
in a good agreement with observed emission characteristics of young
$\gamma$-ray pulsars, including features of the phase resolved
spectra. Second we find model parameters that fit each pulsar
belonging to a group of eight bright pulsars that have a published
phase-resolved spectrum. The $\sigma$ values that best describe each
of the pulsars in this group show an increase with the
spin-down rate $(\dot{E})$ and a decrease with the pulsar age,
expected if pair cascades are providing the magnetospheric
conductivity. Finally, we explore the limits of our analysis and
suggest future directions for improving such models.

\end{abstract}

\section{Introduction}

Pulsars are collapsed cores of massive stars spinning at periods in
the range $\sim 10^{-3} - 10$ sec. They are capable of 
radiating at  almost all spectral wavelengths (from radio to
$\gamma$-rays), although only a small subset have detected 
$\gamma$-ray pulsations. Their emission is highly anisotropic, originating
from particles accelerated along the open magnetic field lines by
intense electric fields induced by their magnetic field rotation.
When pulsars display $\gamma$-ray emission, most of
their luminosity is emitted in $\gsim 100$ MeV $\gamma-$rays. As
such they have been one of the major class of objects studied by the
\textit{Fermi} $\gamma$-ray Space Telescope (\textit{Fermi}), using
the Large Area Telescope (LAT) \citep{LAT}. The release of the
$2^{nd}$ \textit{Fermi} Pulsar catalog (2PC) \citep{2ndFERMIcat},
presents light curves and spectra of 117 $\gamma$-ray pulsars.
Of these, 41 are radio-loud $\gamma$-ray pulsars discovered
through radio timing, 36 are new pulsars discovered through their
$\gamma$-ray pulsations alone, and a surprising 40 are radio-loud
millisecond pulsars. We note that for all the radio-loud
\emph{Fermi} pulsars the radio luminosity is orders of magnitude
smaller than that of $\gamma$-rays. Nonetheless, since the number of radio
photons far exceeds the number of the $\gamma$-ray photons 
radio detection is easier, while 
$\gamma$-ray detection is efficient for the nearer and more
energetic objects \citep{2ndFERMIcat}.

Pulsar radiation has been studied over the years with a variety of
models employing different assumptions about the location and
geometry of emission zones, such as the Polar Cap (PC) \citep{DauHar96},
the Outer Gap (OG) \citep{ChengOG}, \citep{RoYa95}, the Slot Gap
(SG) \citep{Ar83}, \cite{MuHa04}, Separatrix
Layer (SL)\citep{FFbaiSpit} and current sheet \citep{UzdenSpit14,Petri12a,CK10,Kal12c,PULSARGAMMA}. 
Many models, such as SG and OG implement curvature radiation (as in this paper) 
to describe the GeV radiation, while others implement
synchrotron radiation powered by the reconnection in the current sheet \citep{Lyu96, UzdenSpit14}.
The increased sensitivity of the {\em Fermi LAT}, with its first observations of a
simple exponential cutoff of the Vela pulsar spectrum 
\citep{VELA09}, ruled out the PC emission model which predicted a
super-exponential cutoff, due to magnetic pair attenuation near the
neutron star surface.  This result, later confirmed for a number of
other pulsars, established that the pulsar $\gamma$-ray emission
originates in the outer magnetosphere, with the OG and the SG models
better fitting observations. However, none of these outer
magnetosphere emission  models can currently account for the entire
pulsar light curve phenomenology \citep[for example
see][]{RomWat10,Pierb12,PierbiOBS}.  Furthermore, none of these
models are consistent with global magnetosphere properties, such as
current closure.

The recent development of 3D, global, force-free
electrodynamics (FFE) \citep{spitk2006,KalCon09,Petri12b}, Magnetohydrodynamical \citep{Komi06,Tchek13} 
and Particle in Cell \citep{PhiSpi14,Cerutti14,Chen14} pulsar magnetosphere models gave an impetus to the study of pulsar
magnetospheres by providing a more realistic picture of their outer
field geometry than the vacuum model previously assumed. However,
because the ideal MHD FFE regime precludes the existence of
electric fields parallel to the magnetic field (${\bf E_{\parallel}} =
0$) and thus the acceleration of particles and the emission of
radiation, more realistic dissipative MHD magnetosphere models have been
developed \citep{TowardRealistic, Li12}. These models allow for ${\bf
E_{\parallel}} \neq 0$ and therefore can accommodate the  production
of radiation. They are constructed by numerically evolving the time
dependent Maxwell's equations, while at the same time employing a
form of Ohm's law to relate the current density ${\bf J}$ to the
fields. This has the general form
\begin{equation}\label{prescr}
\mathbf{J} = c\rho\dfrac{\mathbf{E}\times\mathbf{B}}{E_{0}^{2}+B^{2}}
+\sigma \mathbf{E_{\parallel}},
\end{equation}
where $\mathbf{J}$ is the current density, $\mathbf{E}$ and
$\mathbf{B}$ are the electric and magnetic field, and $E_{0}$ is
defined by $B_0^2 - E_0^2=\mathbf{B^2} - \mathbf{E^2}$ ,
$E_0B_0=\mathbf{E\cdot B}$, $E_0\geq0$. $E_{0}$ is a term that
prevents the drift velocity and hence the current from becoming
superluminal, while $\sigma$ is a phenomenological conductivity,
used to relate ${\bf E_{\parallel}}$ to the current density ${\bf
J}$, measured in units of the pulsar rotation frequency $\Omega$,
the fundamental frequency in the problem. The macroscopic
spatial $\sigma$ distribution should be consistent with the
underlying microphysics. However, the dependence of the conductivity
on the microphysics is still not understood. Thus, the goal of the recent study 
of global dissipative MHD magnetosphere models is to explore the main properties of
various $\sigma$ values and distributions. In our first studies \citep{Kalap12a, TowardRealistic} a
conductivity that is mainly uniform (in space) and constant in time was assumed. In
\cite{PULSARGAMMA}, we have started to implement  a broad
range of conductivity values along the open magnetic field lines
while the FFE (in reality highly conductive) condition was applied
in the `closed' region.

We have employed these dissipative magnetosphere models to
generate model $\gamma$-ray light-curves \citep{TowardRealistic, PULSARGAMMA} due to curvature radiation (CR).
To this end, we calculate the trajectories and Lorentz factors
$\gamma_L$ of radiating particles, under the influence of both the
accelerating magnetospheric electric fields and CR-reaction. This
approach allowed us to relate the observed pulsar emission
properties to those of the model magnetospheres. Such observables
are the separation (in phase) of the $\gamma-$ray light curve peaks
$\Delta$ and also the lag $\delta$ between the radio emission and
the first peak of the $\gamma-$ray light curve. It was found
\citep{PULSARGAMMA} that matching the model $\delta - \Delta$
distribution with that obtained in the 2PC \citep{2ndFERMIcat} cannot be achieved with a
$\sigma$ constant across the entire magnetosphere but requires
essentially FFE (infinite conductivity, ${\bf
E_{\parallel}}=0$) conditions interior to the light cylinder (LC),
with radius $R_{\rm LC} = c/ \Omega$, and large ($\sigma \gsim 30
\Omega$) but finite conductivity outside the LC giving these models
the nomenclature FIDO (FFE Inside, Dissipative Outside).

In the current paper we expand our study by considering the
FIDO models that were presented in \citep{PULSARGAMMA} by
exploring their energetic and spectral properties under certain
simple assumptions. We attempt to compare the detailed luminosities
and spectra of the dissipative magnetosphere models to those
obtained by observations. To this end, rather than perform a
statistical comparison of the FIDO model with the entire Fermi
population, we decided to compare the phase-averaged and
phase-resolved spectra predicted by the FIDO model with a few very
luminous pulsars that have published phase-resolved spectra. This
being the first comparison of spectra predicted by dissipative
models with data, we chose to focus on individual pulsars by which
we can better understand the model comparison. This
comparison will provide the trends and will reveal the limitations
of this model and its assumptions in describing the
entire high energy pulsar phenomenology depicted by \emph{Fermi}.

In \S 2 we describe the methods employed in our modeling of the
pulsar magnetospheres and the production of the phase resolved
spectra and in \S 3 we present our results. We conclude in \S 4 with
a discussion of their importance of the results and directions
to be followed in the future studies.

\section{Methods}
\begin{figure*}[htbp]
\centering
\includegraphics[width=\textwidth,trim=0 0 0 0]{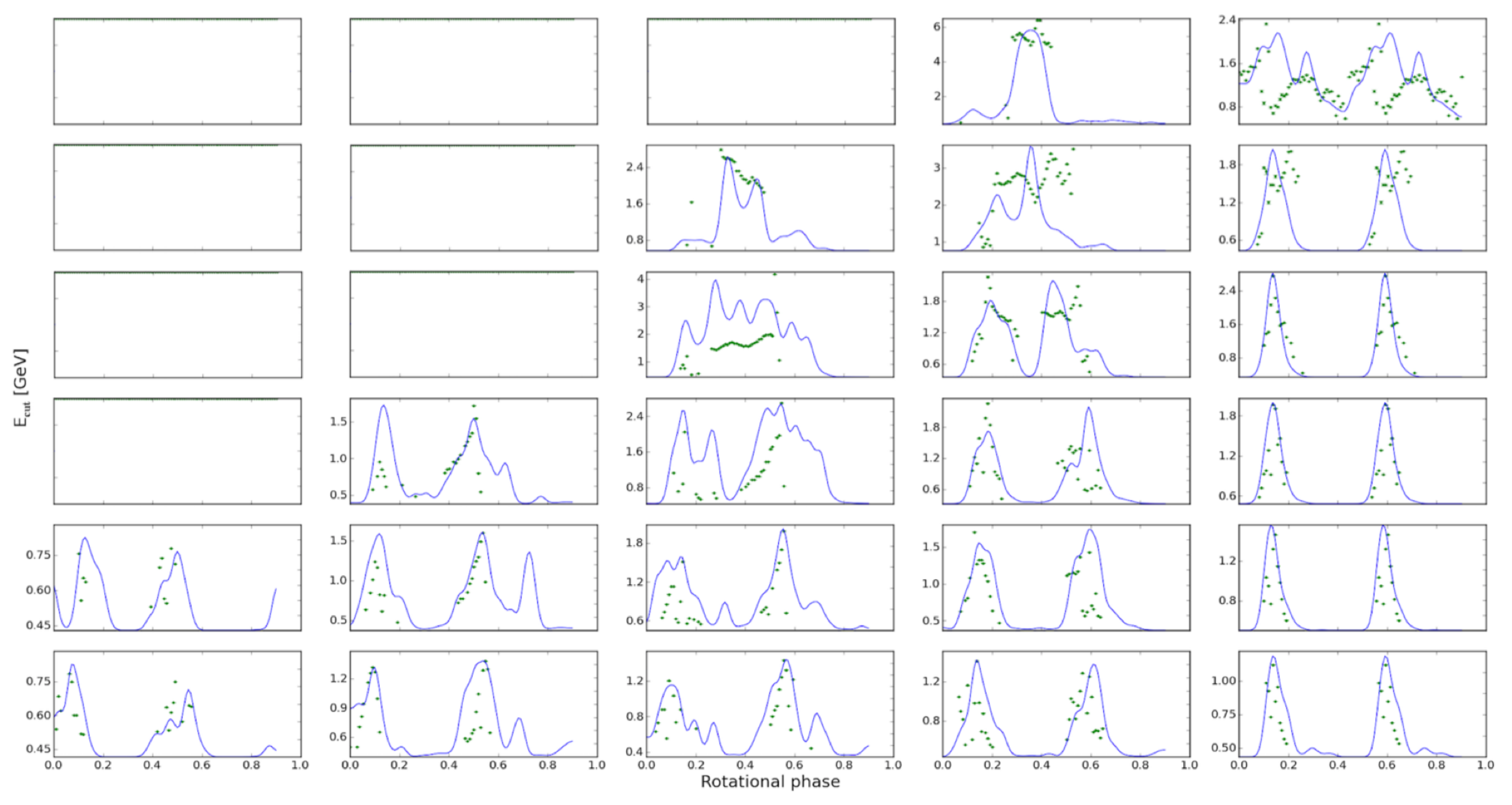}
\caption{Examples of photon light curves, between 100 MeV and 50 GeV,
produced by a FIDO magnetosphere with $P=0.1s$, $B=4\cdot 10^{12}G$
and $\sigma=30 \Omega$. From left to right are different $\zeta$
($30^{\circ}$, $45^{\circ}$, $60^{\circ}$, $75^{\circ}$,
$90^{\circ}$) while from top to the bottom are different
$\alpha$ ($15^{\circ}$, $30^{\circ}$, $45^{\circ}$, $60^{\circ}$,
$75^{\circ}$, $90^{\circ}$). The overlapping green dots are the phase
resolved $E_{cut}$. The \textit{y} axes are rescaled to the relative
maximum of the light curve, the \textit{x} axis is rotation phase, from 0 to 1. The
values on the \textit{y} axis are the values of the phase resolved
$E_{cut}$ in GeV. \label{LCs}}
\end{figure*}

\subsection{The Magnetosphere Structure} \label{magne}
The magnetosphere structure is obtained by numerically solving the time
dependent Maxwell equations
\begin{equation}\label{1MAX}
\dfrac{\partial \mathbf{B}}{\partial t}=-c\nabla\times\mathbf{E}
\end{equation}
\begin{equation}\label{2MAX}
\dfrac{\partial \mathbf{E}}{\partial t}= c\nabla\times\mathbf{B}-4\pi\mathbf{J}
\end{equation}
with a 3D finite difference time domain technique \citep{KalCon09}.
We consider the presence of a dipole magnetic moment
$\boldsymbol\mu$ at the center of the star and that the star itself
is a perfect conductor; then the boundary condition on the stellar
surface for the electric field is the FFE condition for a
corotating magnetosphere:

\begin{equation}\label{FFco}
\mathbf{E}+\mathbf{(\boldsymbol\Omega \times \mathbf{R}) \times B} =0
\end{equation}
The closure of the system requires a prescription for the current
density \textbf{J} in terms of the fields (Eq. \ref{prescr}), as in
\cite{PULSARGAMMA}.  When $\sigma$ is very close to the FFE case
($\sigma\rightarrow\infty, E_{\parallel}\rightarrow 0,
\dfrac{\partial \mathbf{E}}{\sigma\partial t}\rightarrow 0$), taking
the parallel component of Eq. \ref{2MAX} it is possible to
approximate the electric field parallel to the magnetic field as
\cite{PULSARGAMMA}
\begin{equation}\label{esig}
E_{\parallel}=\dfrac{c|(\nabla \times \mathbf{B})_{\parallel}|_{(\rm FFE)}}{4\pi\sigma}
\end{equation}
where the FFE indication implies the corresponding
FFE value.\footnote{This equation
comes from a simplified Maxwell equation in which the term
$\sfrac{\partial E_{\parallel}}{\sigma\partial t}$ is neglected.}

As in \cite{PULSARGAMMA}, we start with a FFE global
magnetosphere model implementing the FFE current density $J$
prescription \citep{G1999}
\begin{equation}\label{IFFEJ}
    \mathbf{J}=c\rho\frac{\mathbf{E}\times\mathbf{B}}{B^2}+
    \frac{c}{4\pi}\frac{\mathbf{B\cdot \nabla\times B - E\cdot \nabla\times
    E}}{B^2}\mathbf{B}
\end{equation}
%
%
and then compute $E_{\parallel}$ only outside the LC, located at
radius $R_{\rm LC} = c/ \Omega$, using the approximation in Eq. \ref{esig}
of the prescription of Eq. \ref{prescr} for $\sigma = 0.3, 1, 3, 5, 10, 30,
60 \Omega$ ($\sigma = 30 \Omega$ was used for the original FIDO
model). We did not test higher $\sigma$ values since for 
these the corresponding cutoff energies $(E_{\rm cut})$ and total
$\gamma$-ray luminosities ($L_{\gamma}$) are well below the ones
observed by \emph{Fermi}. In our FFE simulations we use a stellar
radius $r_{\star}= 0.2R_{\rm LC}$. Our computational domains extend
up to $5R_{\rm LC}$ and the spatial resolution (i.e. the grid cell
size) is $0.01R_{\rm LC}$. Each simulation has been evolved for ~2
stellar rotations.

Here we would like to note that ideal (of infinite
resolution) FFE simulations would lead to infinitesimally thin
current sheets and hence to infinitely large $\nabla \times
\mathbf{B}_{{\parallel}_{(\rm FFE)}}$ and $E_{\parallel}$ values
lying within infinitesimally small volumes. However, in
dissipative solutions (finite $\sigma$) the thickness of the current
sheet is expected to be of the order of $c/\sigma$. In our FFE
simulations the resolution is finite ($0.01R_{LC}$) and the
corresponding current sheet is resolved within a few grid cells.
This behavior mimics the finite thickness of the current sheet in
dissipative solutions. This ``artificial" FFE thickness of the
current sheet is similar to that corresponding to ``real"
dissipative solutions with $\sigma=30 \Omega$. Our base FIDO solution is
the one with $\sigma=30 \Omega$ presented in Kalapotharakos et al. 2014. We
have checked that the results corresponding to FFE solutions with a
spatial distribution of $E_{\parallel}$ based on Eq.~5 (assuming
$\sigma=30 \Omega$) and those of ``true" dissipative solutions (using
explicitly prescription \ref{prescr}) with uniform $\sigma=30 \Omega$
provide very similar results.

We  also checked the validity of this approximation by
simulating a few FIDO magnetospheres for $\sigma=0.3, 5 \Omega$ and $10 \Omega$
and various magnetic moment inclination angles with
respect to the rotational axis, $\alpha$ (fixing the period, P, and
B); we found that especially for $\alpha<45^{\circ}$ and
for $\sigma\leq5 \Omega$ the field structure deviates substantially from
the FFE field structure on which our approximation is
based. This is primarily because $\sigma$ has been assumed to be
uniform outside the light cylinder. However, the important region
for the particle acceleration in the FIDO model is near the
equatorial current sheet outside the light cylinder
(\citealt{PULSARGAMMA}). In the case of small $\sigma$ values, a
more accurate treatment is needed (e.g. application of the smaller
$\sigma$ values only near the equatorial current sheet). This study
is beyond the scope of the current paper and will be addressed in a
future work (see \S4). In the rest of the paper we consider
the FFE numerical models employing an $E_{\parallel}$ that rescales
according to Eq.~\eqref{esig}. This treatment, though simplistic,
allows the exploration of the main trends and correlations among the
various parameters taking into account a broad range of properties
of the observed phenomenology (e.g. light curves, spectra).

\subsection{Simulation of the emission}

To simulate the $\gamma$-ray emission of the model pulsar
magnetospheres we neglect the inverse Compton (IC) and synchrotron
radiation (SR) contribution and consider only curvature radiation
CR, \citep{Jackson}.  In some models, where particles are
energized by reconnection in current sheet \citep{UzdenSpit14}, SR is
produced by high temperature particles up to 100 GeV. In the type of
model we are considering here, where particles are energized only by
the induced electric field parallel to the magnetic field, the
particles can emit SR only if they acquire some pitch angles. In
such ``gap" models \citep{Harding08,Tang08} SR does not give
photons more energetic than 100 MeV. IC requires a high-altitude
non-thermal X-ray emission component that is present only for a few
\textit{Fermi} pulsars such as the Crab.  To save computational time
in calculating the emission, a random subset of $1.5\cdot 10^{6}$
initial particle orbit positions were selected on the pulsar polar
cap. Their orbits were integrated under the influence of the local
parallel electric field $E_{\parallel}$ and losses due to curvature
radiation to compute their Lorentz factor $\gamma_L$ as a function of
position, in the non-rotating inertial frame. Then we selected
randomly $5\cdot 10^{10}$ particles positions along these
trajectories between the star surface and 2.5 times the light
cylinder radius $R_{LC}$, where we computed the spectrum of the
locally emitted CR and its direction $(\zeta,
\phi)$, taking into account the time delays and assuming that the CR
photons are emitted in the direction of particle motion ($\zeta$ is
the observer's inclination relative to the pulsar rotation axis and
$\phi$ the
azimuth). 
For each particle we calculated the number of photons emitted per
second by CR and stored them in a 3D matrix: 100 bins of resolution
for the azimuth $\phi$ (between $0^{\circ}$ and $360^{\circ}$), 180
bins of resolution for $\zeta$ (between $0^{\circ}$ and
$180^{\circ}$) and 114 energy bins equally spaced logarithmically
between 0.01 and 50 GeV.

We calculated the luminosity by rescaling the area of the polar cap
according to the period of the star and to a fixed radius of 10 km
(in the simulation of the FIDO model the radius was at 0.2$R_{LC}$)
and the flux of particles from the surface with the Goldreich-Julian
(GJ) density $\rho_{\rm GJ}$ (\cite{GJ}), assuming the
polar cap is small enough to consider the GJ charge density
constant\footnote{This condition is broken in millisecond pulsars
because their polar cap size is a significant fraction of their
radius.}. The adopted $\rho_{\rm GJ}$ flux assumes that
the multiplicity of the accelerated particles in the regions with
high $E_{\parallel}$ is small ($\sim1$).  
This assumption is reasonable given that much higher multiplicity 
leads to screening of the $E_{\parallel}$ over distances short enough to  
prevent fluxes $ \gg \rho_{\rm GJ}$ of accelerated particles.
This 3D grid allows us to calculate sky maps, light curves and
phased resolved spectra. The sky maps are produced by summing up
over the photon energy and plotting the resulting intensity in
photons/s in ($\zeta, \phi$) coordinates. Summing over energy for a
particular viewing angle $\zeta$ of the sky map produces a light
curve. Phase-averaged spectra can be obtained by averaging the
photon distribution in energy over $\phi$ for a chosen $\zeta$.
Finally, phase-resolved spectra are obtained by collecting the
spectra in each ($\zeta, \phi)-$bin. We applied this
procedure to pulsar magnetospheres with a number of inclination
angles $\alpha$ between the rotation and magnetic axes ($\alpha =
15^{\circ}, 30^{\circ}, 45^{\circ}, 60^{\circ}, 75^{\circ},
90^{\circ}$), different rotation periods ($0.01 s, 0.03 s, 0.1 s,
0.3 s, 1 s$) and different magnetic fields at the stellar surface
($5\cdot 10^{11}$G, $10^{12}$G, $2\cdot 10^{12}$G, $4\cdot
10^{12}$G, $8\cdot 10^{12}$G, $3\cdot 10^{13}$G) assuming each time
all the different conductivity values
$(\sigma=0.3\Omega,1\Omega,3\Omega,5\Omega,10\Omega,30\Omega,60\Omega)$
mentioned in the previous section. Thus, we tested
$(6\times7\times5\times6=1260)$ different magnetosphere
configurations.

The sky maps and light curves shown in \cite{PULSARGAMMA} were
computed for the bolometric luminosity, while here we computed sky
maps and light curves in photons/s in the energy band $0.1 - 50$ GeV
in order to compare with observed {\it Fermi} light curves and
spectra.  Our light curves in Figure 1 show some extra secondary
peaks that do not appear in the bolometric light curves, due to
relative spectral differences with phase. We generated sky maps and
light curves in bolometric luminosity and they are very similar to
those of \cite{PULSARGAMMA} except for small differences in the
relative heights of the light curve peaks (the maximum difference is
$\lsim 20\%$).  This kind of difference is due to differences in
sampling and/or binning. In addition, we smoothed the profiles (with
a gaussian smoothing\footnote{$y=\exp [-\sfrac{x^{2}}{\Theta^{2}}]$}
with $\Theta = 3.6^{\circ}$) to reduce the numerical noise.

\subsection{Comparison with the {\it Fermi} pulsars}
We used the data from the 2PC, gathered with three years of
observations acquired by the LAT on the \textit{Fermi} satellite,
except for the phase-resolved spectra that have appeared in
different papers (\cite{Megan}, \cite{threePul}, \cite{J1836}).
These different works use data collected over different amounts of
time (generally less than the 3 years of data used for 2PC) but they
all analyze photons in the same energy range (between 100 MeV and 50
GeV like in the 2PC).

\begin{figure}[htbp]
\centering
\includegraphics[width=9.0cm,trim=10 0 10 0]{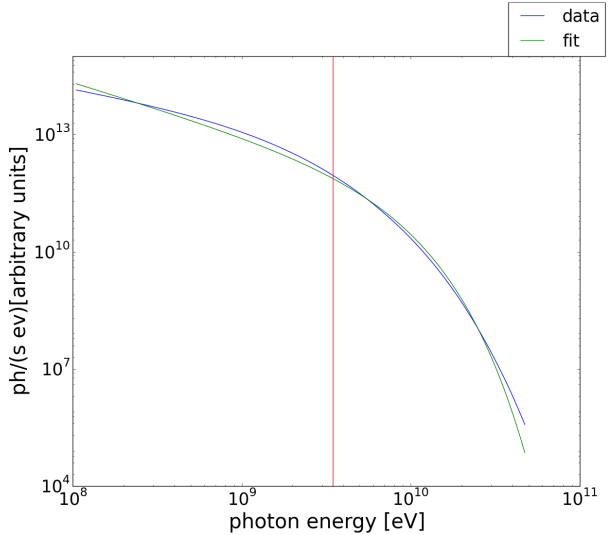}
\caption{An example of a phase-averaged spectrum produced by the model (blue) 
and fitted to a power law with exponential cutoff (green). The red
vertical line is the fit value of $E_{cut}$. The \textit{x} axis 
is photon energy in eV, the \textit{y} axis is a quantity
proportional to the amplitude. \label{specPROVA}}
\end{figure}

We study some particularly luminous $\gamma$-ray pulsars, in
particular the ones that have a  published phase resolved spectrum.
This set of eight pulsars contains: J0007+7303 (CTA1 pulsar),
J0534+2200 (Crab pulsar), J0633+1746 (Geminga pulsar), J0835-4510
(Vela pulsar), J1057-5226, J1709-4429, J1836+5925, J1952+3252. We
identified a candidate light curve and its associated phase-resolved
spectrum produced by our model for a combination of $\alpha$,
$\zeta$, $\sigma$ that best describe those of each pulsar. We made a
first selection using the number of peaks in the light curve.  The
sample of model light curves were generated at different viewing
angles $\zeta$ (equivalent of the latitude) between $5^{\circ}$ and
$90^{\circ}$ with a step of $5^{\circ}$. These 18 light curves were
generated for each inclination angle and for $\sigma=1 \Omega,5 \Omega,30 \Omega$. This
grid was fine enough to resolve the single/double peaked attribute
of the light curves, identifying the best values of
$(\alpha,\zeta)$.  We then used simulated magnetosphere models for
the whole range of $\sigma$ noted above, selecting the closest P and
B values to those of the real pulsars. For each combination of
$\alpha$ and $\zeta$ selected we matched the cutoff energy $E_{cut}$
from the phase averaged spectra (an example in Figure
\ref{specPROVA}) using the same form\footnote{This expression is the
same as used in \citep{2ndFERMIcat} with \textit{b}=1.} as was used
to fit the pulsar spectra of the 2PC \citep{2ndFERMIcat}
\begin{equation}\label{spectra}
\frac{dN}{dE}=K\left(\frac{E}{E_{0}}\right)^{-\Gamma}\exp\left(-\frac{E}{E_{cut}}\right)
\end{equation}

We looked for a candidate (considering the whole range of $\sigma$
values) with an $E_{cut}$ that was within a factor of 2 of that
observed. We then matched the separation between the gamma-ray peaks
(with a tolerance of 0.08 in phase, where 1.0 is the period), the
evolution of light curves with energy predicted by the model, and
the flux, in comparison with that observed in the 2PC.  
The distances used are those reported in the 2PC.  Considering the
maximum error in the distance measures and the maximum variation in
the flux of the simulated light curves with respect to $\alpha$, and the 
assumed multiplicity of 1 in the particle flux, we
accept a flux within a factor of 10 of the observed value. These
were the primary features we considered in selecting candidate light
curves and spectra. The other features we considered were the
``half-power width" of the light curve peaks and the trend of the
phase-resolved $E_{cut}$. We did not give as much importance to the
spectral index $\Gamma$ because it is more strongly influenced by
possible contributions from SR and IC emission. We note that the
discrepancies in the model flux that we would have obtained
considering the observed spectral index ($\Gamma$) value in the
phase averaged spectrum were always less than 15\%.

\section{Results}
\subsection{General properties of the model}

As was reported in \cite{PULSARGAMMA} the FIDO model has the
interesting property of reproducing the correlation between $\delta$
and $\Delta$ presented in the 2PC \citep{2ndFERMIcat}. The simulated
light curves in Figure \ref{LCs} look similar to the observed light
curves, except for some additional smaller peaks. In Figure
\ref{skymaps} of the Appendix we show the luminosity skymaps of a
FIDO magnetosphere with $P=0.1s$, $B=4\cdot 10^{12}G$ and
$\sigma=5 \Omega$;  we also plotted the trends in flux and $E_{cut}$ with
$\alpha$, $\zeta$ and $\sigma$. We notice that the flux and the
$E_{cut}$ generally increase with $\zeta$ and decrease with
$\sigma$, for fixed $\alpha$. They also increase with decreasing
\textit{P} and increasing \textit{B}. The physical cause of these
trends are that increasing $\zeta$ more closely approaches the more
energetic main caustic (as we can see in the Appendix in Figure
\ref{skymaps}). Increasing $\sigma$ decreases the value of
$E_{\parallel}$, the maximum value of the particle Lorentz
factor $\gamma_L$ (as we noted in subsection \ref{magne}) and hence
the value of $E_{cut}$.
Increasing \textit{B} also increases $E_{\parallel}$ (fixing
$\sigma$), while decreasing \textit{P} reduces the LC radius and 
thus on the one hand decreases the radius of curvature of the
particle trajectories (in absolute units) and on the other hand
increases the absolute $E_{\parallel}$ values (the closer to the
star the emitting region is the higher the corresponding field
values are). These effects, in general, increase the $\gamma_{L}$
values, the corresponding photon energies ($E_{cut}$), and the total
power emitted. Increasing $\alpha$ results in a decrease of
$E_{cut}$ and of the flux: this is due to both $E_{\parallel}$ being
smaller and the GJ charge density (\cite{GJ}) at the surface of the
polar-cap decreasing with $\alpha$, reaching 0 at
$\alpha=90^{\circ}$.\footnote{For $\alpha=90^{\circ}$ we substituted
the GJ density at $85^{\circ}$ to avoid having zero charge density
for this case.}  

In Figure \ref{lumSEL}, of the Appendix we show the
dependence of the luminosity of our models with the pulsar period
and magnetic field (red dots) as well as the observations of 2PC
(blue dots). In Figure \ref{cuVA} we show the dependence of
$E_{cut}$ for a pulsar with $P = 0.1$ sec and $B = 2 \times 10^{12}$
G as a function of $\alpha, \zeta, \sigma$ while in Fig.~\ref{fluVA}
the dependence of flux of the same pulsar on the same parameters
assuming a distance of 1 kpc. Generally, the model $E_{cut}$ 
values for $\sigma \ge 3 \Omega$ match the observed values. while for 
$\sigma < 3 \Omega$ the model $E_{cut}$ values are larger than observed.
The model flux values for all $\sigma$ are within the observed range,
except for the largest $\sigma$ values at the smallest $\alpha$ and $\zeta$.
Although the model fluxes are about a factor of ten 
below the highest observed fluxes at the largest $\alpha$, even for the lowest 
$\sigma$ values, Fig.~\ref{fluVA} is only plotted for one value of $P$ and $B$
and does not show model fluxes for the full range of these values in the 
observed population.

Another feature we investigate is the general trend of the phase
resolved spectra. From the phase
resolved spectra that are available it seems that the $E_{cut}$ is
always higher in the $2^{nd}$ light curve peak or in the right part
of a broad single peaked light curves. This behavior is followed by 
$\sim 55\%$ of our model light curves, $23\%$  do not
show a significant difference in $E_{cut}$ between the two peaks,
while $22\%$ exhibit the opposite behavior. So, according to the
FIDO model, obtaining phase-resolved spectra from more pulsars
should show some that do not follow the trend of higher $E_{cut}$ in
the second peak.

In a few of the magnetosphere models, the luminosity exceeds
$\dot{E}$. Since this cannot happen physically, the problem is
linked to the following cause. 
The approximation of Eq.~\ref{esig} forces the
$E_{\parallel}$ to change linearly with respect to $\sigma$.
Nonetheless, we found through full simulations using low $\sigma$
that this process is not exactly linear and the $E_{\parallel}$
saturates for low $\sigma$ values. Thus, whenever the
$E_{\parallel}$ is higher than it should have been this can
contribute to an overestimation of the luminosity. Finally, this
effect is probably due to the overestimation of the distances of the
given pulsars.

\subsection{Model $\gamma$-ray pulsar identification}
First we performed a \textit{blind-search} to find the parameters
giving light curves and spectra that best match those of the Fermi
pulsars, disregarding any constraints from other wavelengths.
We note that on the one hand the light-curve form is the
most important feature for the determination of each pulsar's
$\alpha$ value and the Earth's $\zeta$ value. On the other hand 
the $E_{\rm cut}$ and $L_{\gamma}$ values determine the
corresponding $\sigma$ value. We found from one to three candidates
per pulsar. But of the 8 pulsars for which we found model candidates
with this \textit{blind-search}, two did not have a good match with
other wavelength constraints (taken from \cite{PierbiOBS}). Looking
for candidates again in the zone indicated by these constraints we
found one possible candidate per pulsar, with a less-than-perfect
light curve match but still reasonable. This fact shows that our
model is not always able to identify in an unambiguous way a
candidate that satisfies all our criteria simultaneously. We report
the geometry and conductivity information of our candidates in Table
\ref{candiT} while the fluxes and the $E_{cut}$ are shown in Table
\ref{candiEF}.
\begin{table}[h]
\begin{center}
\caption{FIDO Pulsar Candidates - Geometry.\label{candiT}}
\begin{tabular}{ccccccc}
\tableline
Name & $\alpha$\footnote{\cite{PierbiOBS}\label{footnotePIERBI}}[$^{\circ}$]  & $\zeta$\footref{footnotePIERBI}[$^{\circ}$] & $\alpha$[$^{\circ}$] & $\zeta$[$^{\circ}$] & $\sigma [\Omega]$ & Radio\\
\tableline
J0007+7303 & -- & -- & 30 & 73 & 10 & \textit{quiet}\\
J0534+2200 & -- & 61 & 75   & 63 & 30 & \textit{caustic}\\
J0633+1746 & -- & $>60$ & 45    & 87 & 3 & \textit{quiet}\\
J0835-4510 & 70 & 64 & 60   & 50 & 10 & \textit{loud}\\
J1057-5226 & -- & -- & 45   & 61 & 30 & \textit{loud}\\
J1709-4429 & -- & 53 & 45 & 55 & 10 & \textit{loud}\\
J1836+5925 & -- & -- & 90   & 40 & 1 & \textit{quiet}\\
J1952+3252 & -- & -- & 75   & 85 & 10 & \textit{loud}\\
\tableline
\end{tabular}
\end{center}
\end{table}

\begin{table}[h]
\begin{center}
\caption{FIDO Pulsar Candidates - Energetics\footnote{$E_{cut}$ are in GeV, fluxes in $10^{-10}\,\rm erg\,s^{-1} cm^{-2}$}}
\label{candiEF}
\begin{tabular}{cccccccc}
\tableline
Name & $E_{cut}^{obs}$ & $\Gamma^{obs}$ & $F_{\gamma}^{obs}$ & $E_{cut}$ & $\Gamma$ & $F_{\gamma}$ & $\sigma [\Omega]$\\
\tableline
    J0007+7303 & 4.7 &1.4& 4.0 & 5.0 &0.9& 0.8 & 10\\
    J0534+2200 & 4.2 &1.9& 13 & 3.4  &1.3& 16 & 30\\
    J0633+1746 & 2.2 &1.2& 42 & 3.4  &1.2& 1.5 & 3\\
    J0835-4510 & 3.0 &1.5& 91 & 4.1 &1.4& 19 & 10\\
    J1057-5226 & 1.4 &1.0& 3.0 & 1.3 &2.0& 1.5 & 30\\
    J1709-4429 & 4.2 &1.6& 14 & 4.2 &1.1& 1.1 & 10\\
    J1836+5925 & 2.0 &1.2& 6.0 & 3.0 &1.1& 0.2 & 1\\
    J1952+3252 & 2.5 &1.5& 1.4 & 2.2 &1.4& 0.7 & 10\\
\tableline
\end{tabular}
\end{center}

\end{table}

\begin{figure}
\centering
\includegraphics[width=9.0cm,trim=0 0 0 0]{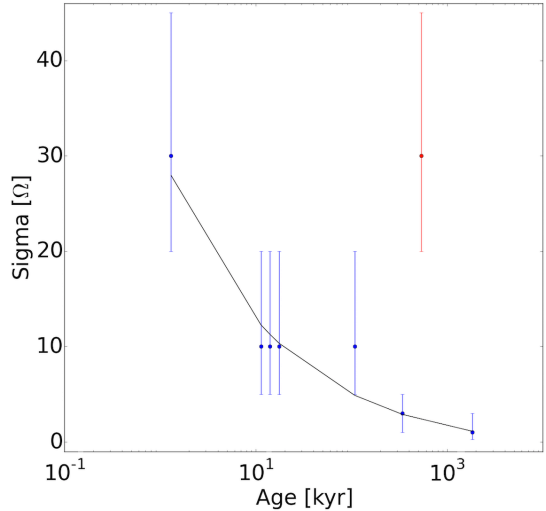}
\caption{The conductivity, $\sigma$, for the best candidates as a
function of pulsar characteristic age. The pulsars used in the fit are in blue;
the one in red is J1057-5226 and was excluded from the fit; the
solid black line is the power law found by the fit routine (see Eq.
\ref{PWLsa}) \label{SAPOW}.}
\end{figure}

\begin{figure}
\centering
\includegraphics[width=9.0cm,trim=0 0 0 0]{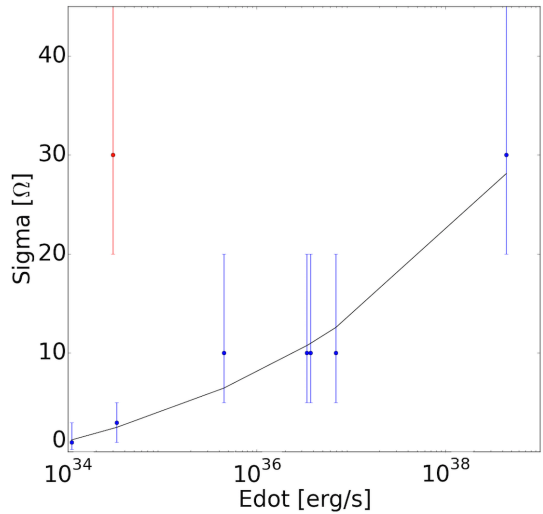}
\caption{The conductivity, $\sigma$, for the best candidates as a
function of pulsar spin down luminosity $\dot{E}$. The pulsars used
in the fit are in blue; the one in red is J1057-5226 and was excluded from
the fit; the solid black line is the power law found by the fit
routine (see Eq.~\ref{PWLse})\label{SEPOW}.}
\end{figure}

We find a correlation between the $\sigma$ of the candidates and
their $\dot{E}$ and the age estimate obtained through the period and period derivative
(excluding the candidate for J1057-5226, since it seems to be an
outlier). We fit these two correlations with a power law,
using the large intervals we had on the $\sigma$ grid as errors:
\begin{equation}\label{PWL}
y = a + b \cdot x^{c}
\end{equation}
We obtained for the relation $\sigma -$ age:
\begin{equation}\label{PWLsa}
a=-1\pm3, b=32\pm23, c=-0.36\pm0.01
\end{equation}
While for the relation $\sigma - \dot{E}$:
\begin{eqnarray}\label{PWLse}
& a=-6\pm30,  b=(7.80\pm 0.02)\times 10^{-5},  \\
& c=0.146\pm0.004 \nonumber
\end{eqnarray}
both with $\chi^{2}_{red}\simeq0.2$, where $\chi^{2}_{red}$ is the reduced 
$\chi^{2}$, that highlight the large error
bar assumed. These relations are shown in Figures \ref{SAPOW},
\ref{SEPOW}.
These two correlations are expected if higher $\sigma$ results from
more efficient screening of $E_{\parallel}$ by pairs that are
produced in greater numbers by younger, more energetic pulsars
\citep{DauHar82,MuHa11}. Our 8 candidates do not
reproduce all of the observed characteristics of these pulsars: two
had a bit smaller flux than observed (within a factor of 15,
J1709-4429, and 40, J0633+1746, instead of 10), two of them do not
reproduce the observed bridge emission in the light curve
(J0633+1746) or the off-peak emission present in the observed light
curve (J1836+5925). For the candidate of radio quiet pulsar Geminga
(J0633+1746)  having a 0.5 peak separation, we interchanged the
model peaks in order to match the observed relative peak heights,
which we are free to do since our model light curve has no bridge
emission.   The J0007+7303 candidate has a smaller peak width than
observed (the tolerance is $\pm0.06$) and two candidates have phase
resolved spectra that do not show the observed trends (J0007+7303,
J1836+5925).
 In the Appendix, we show observed and model light curves and phase-resolved spectra for three of the pulsars, Vela, Crab and J1952+3252.

\section{Conclusion}
We have computed light curves, luminosities, phase-resolved and
phase-averaged spectra for a large number of FIDO models (which were
shown to reproduce the $\delta - \Delta$ distribution of 2PC)
for different values of $\alpha, P$, $B$, $\sigma$
and $\zeta$. All of these emission characteristics of the model
were compared with light curves, phase-averaged and phase-resolved
spectra of eight bright pulsars studied with {\it Fermi}. In
particular we explored whether the common trends found in the eight
published phase-resolved spectra, such as a higher $E_{cut}$ in the
second peak, are absolute predictions of the FIDO model. We found
that this is in fact a common trend, but not absolute since the
reverse trend (higher $E_{cut}$ in the first peak) or similar
$E_{cut}$ in both peaks, is also predicted for a subset of
parameters. It would therefore be of interest to produce
phase-resolved spectra for a larger number of pulsars.

This work shows that the FIDO dissipative model merits further
exploration, particularly with respect to the correlations seen in
Figures \ref{SAPOW} and \ref{SEPOW}.  If these correlations are
confirmed, it would strengthen the expected connection between
screening of $E_{\parallel}$ by efficient pair production and high
conductivity.

The major limitations of the present model are the assumption of
FFE scaling of $E_{\parallel}$ with conductivity outside
the light cylinder, and the assumption that the magnetosphere inside
the light cylinder is completely FFE. 
The choice of spatial variation of the conductivity
(infinite inside and finite outside the LC) was imposed by the need
to reproduce the pulsar $\delta$-$\Delta$ (radio lag - peak
separation) correlations. Consistency of the models with these
correlations requires no (or very little) emission from within the
LC, hence the FFE approximation for this region. Clearly, emission
of radiation requires finite conductivity and non-zero
$E_{\parallel}$ somewhere in the magnetosphere, necessarily outside
the LC. A more realistic distribution of $\sigma$ likely requires
its variation with distance along the field lines and from one line
to another. This variation prescription should be eventually
consistent with the related microphysical mechanisms (the details of
which are not well understood, at present at least). Nonetheless,
the overall distribution cannot be much different than that proposed
in Kalapotharakos 2014, else it would be inconsistent with the
$\gamma$-ray light curve correlations.

We have also explored the validity of the FIDO model by
making some test simulations with low $\sigma$ outside the light
cylinder, still assuming FFE conditions inside. It was
found that the $E_\parallel$ in these simulations does not drop
linearly with $\sigma$ for $\sigma < 5 \Omega$ and $\alpha < 45^\circ$,
primarily because there is significant departure from FFE field
structure. Therefore, the fluxes and luminosities are somewhat
overestimated for the candidates with smaller $\sigma$.  However, we
have also begun to explore models with low $\sigma$ only near the
open field boundary, extending outside the light cylinder along the
current sheet.  This distribution is more realistic since screening
by pair cascades is expected for most of the interior field lines.
In this case, the FFE field field structure is present even
for lower $\sigma$ values, indicating that the FFE scaling
of $E_\parallel$ is more valid for a gap-like distribution of
$\sigma$.

After having explored the possibilities of FIDO-like models, we will
continue necessary improvements. First, the magnetosphere
simulations should be calculated exactly and not assuming the linear
behavior of Eq. \ref{esig}. Second, an increase in the grid of
simulated values is warranted, in particular for the $\alpha$
values, because it is the primary source of error in the
determination of the simulated flux.  Producing a large atlas with
more light curves would better track rapid changes through the grid
values.  A further step would be to try to understand the origin of
the $\sigma$ parameter and its spatial variation, in
particular the underlying pair production process at different
sites, such as the polar caps \citep{PPTim}, outer gaps
\citep{ChengOG} or the current sheet.

\bibliographystyle{apj}
\bibliography{pulsargamma}
\appendix
\begin{figure*}[htbp]
\centering
\includegraphics[width=\textwidth,trim=0 0 0 0]{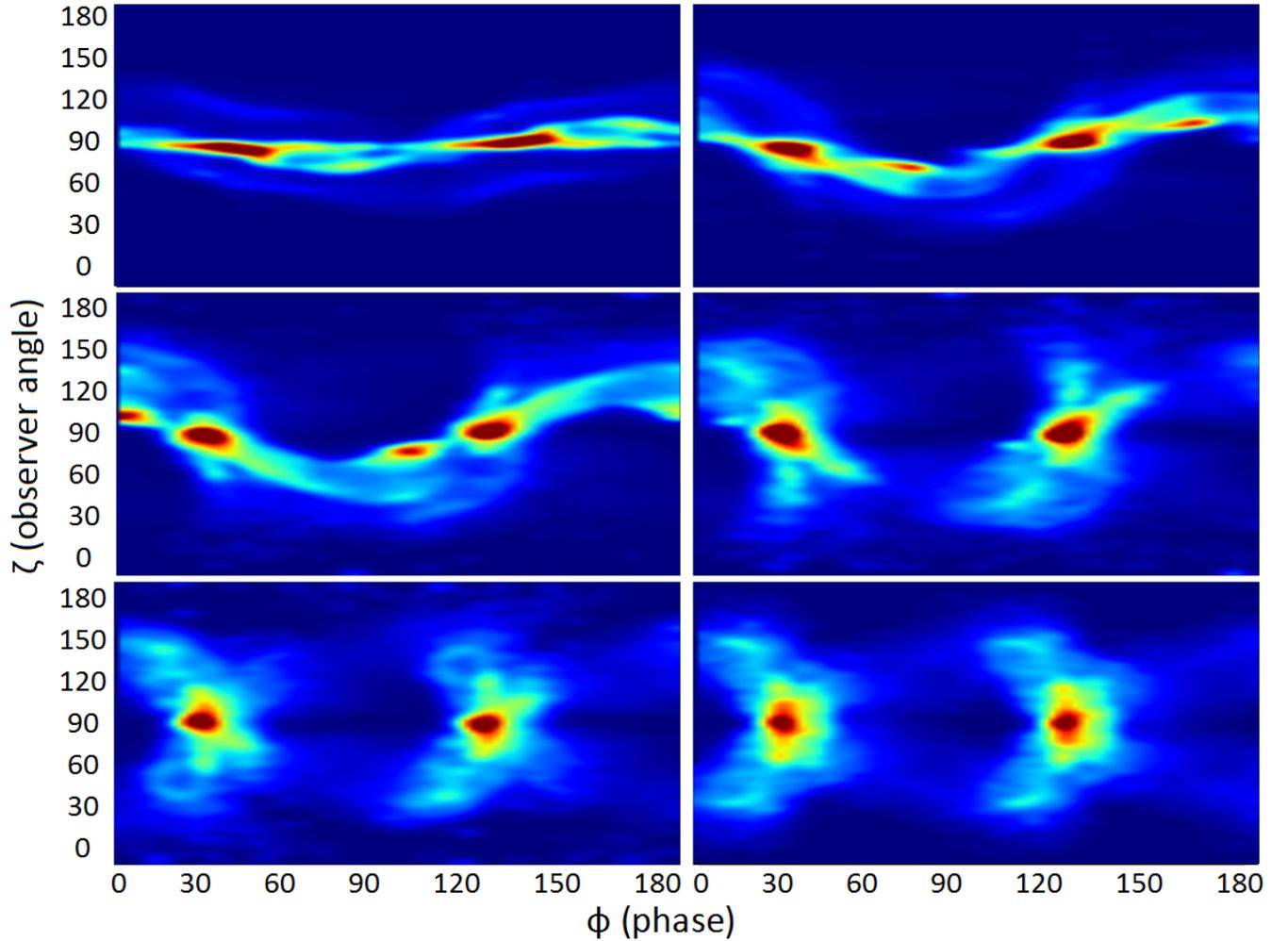}
\caption{Skymaps produced by a FIDO magnetosphere with $P=0.1s$,
$B=4\cdot 10^{12}G$ and $\sigma=5 \Omega$ for the energy range $0.01 - 50$ GeV. 
Top panels, left to right, are $\alpha=15^\circ, 30^\circ$, middle panels are 
$\alpha= 45^\circ,60^\circ$, and bottom panels are $\alpha=75^\circ,90^\circ$.  
The scale is normalized linearly to 99 percentile of the relative
maximum. The \textit{x} axis is rotation phase $\phi$, in degrees; the
\textit{y} axis is observer angle $\zeta$, in degrees, with respect to the
rotation axis. \label{skymaps}}
\end{figure*}

\begin{figure*}
\centering
\includegraphics[width=\textwidth,trim=0 0 0 0]{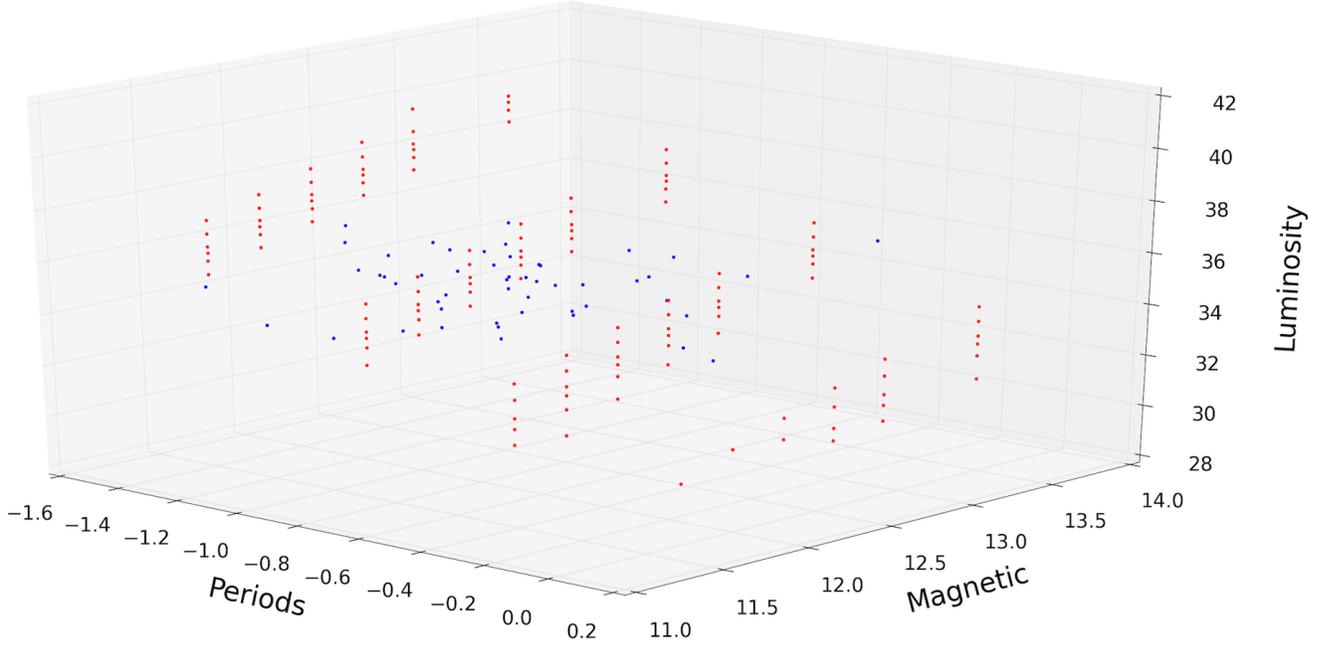}
\caption{FIDO model luminosity over the set of
parameters $(\alpha,\sigma,P,B)$ (red) and the luminosity of the
pulsars in the 2PC (blue), obtained by assuming the flux correction
factor, $f_{\Omega} = 1$  \citep{2ndFERMIcat}. The horizontal axes
are the $log_{10}$ of periods and the $log_{10}$ of magnetic field
values at the pole: the first are measured in \textit{s}, the second
in \textit{G}. The vertical axis is the $log_{10}$ of luminosity 
measured in erg/s. \label{lumSEL}}
\end{figure*}
\begin{figure*}
\centering
\includegraphics[angle=0,width=\textwidth,trim=0 0 0 0]{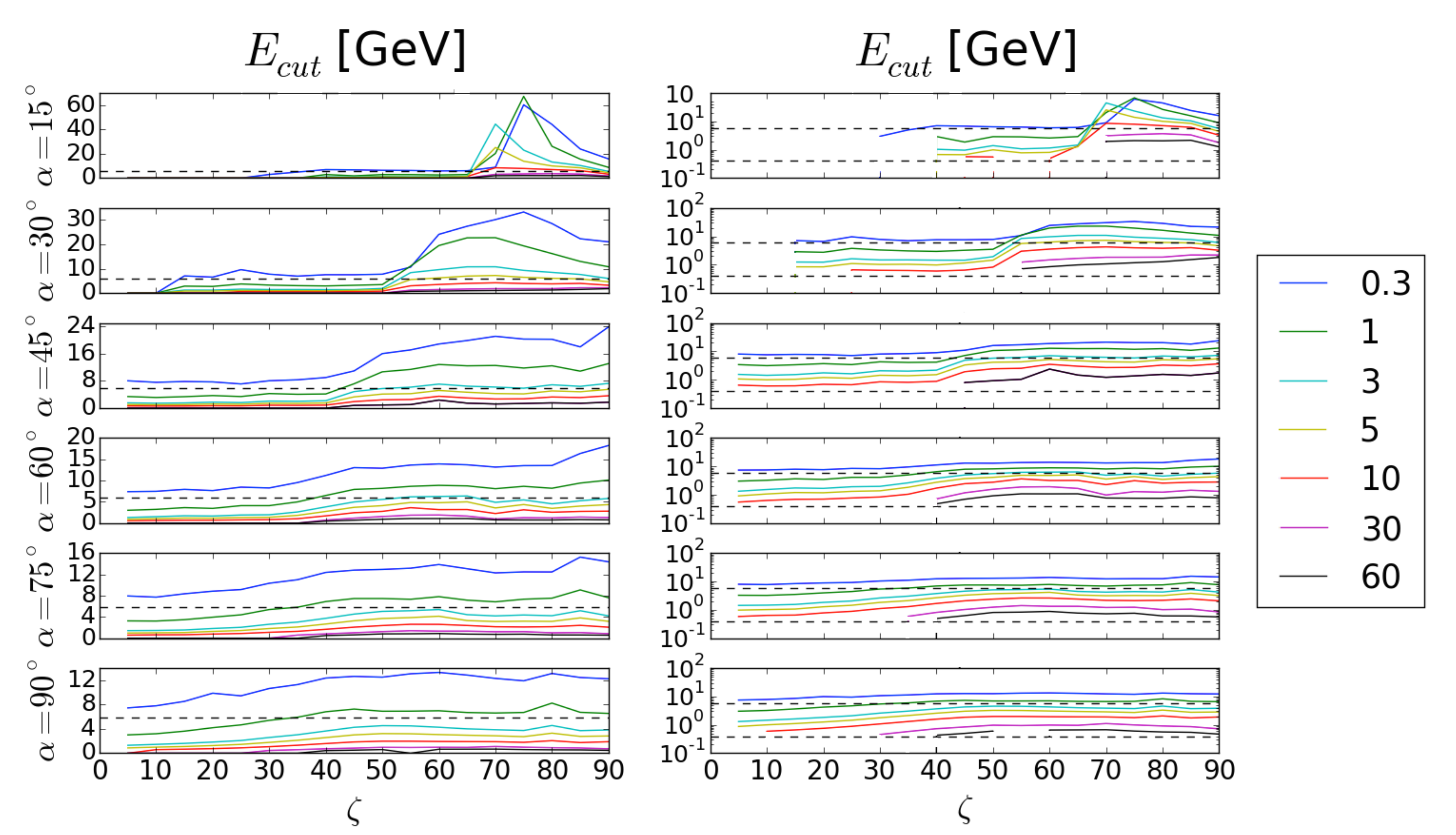}
\caption{The evolution of $E_{cut}$ for a pulsar with $P = 0.1s$
and $B=2\cdot10^{12}G$ for different $(\alpha,\zeta,\sigma)$. 
On the left there is a linear scale, on the right logarithmic. The
six panels are the different $\alpha$ starting from the top to the
bottom: $15^\circ, 30^\circ, 45^\circ, 60^\circ, 75^\circ, 90^\circ$. Line colors denote 
different $\sigma$ (in units of $ \Omega$), according to the legend at the side of the plot.  
The x axis is $\zeta$ from $0^{\circ}$ to
$90^{\circ}$, the y axis is the $E_{cut}$ in
GeV. The dashed lines are the maximum and minimum $E_{cut}$ measured for the Fermi young $\gamma$-ray pulsars. 
In the linear scale panel the minimum $E_{cut}$ is not plotted. \label{cuVA}}
\end{figure*}
\begin{figure*}
\centering
\includegraphics[angle=0,width=\textwidth,trim=0 0 0 0]{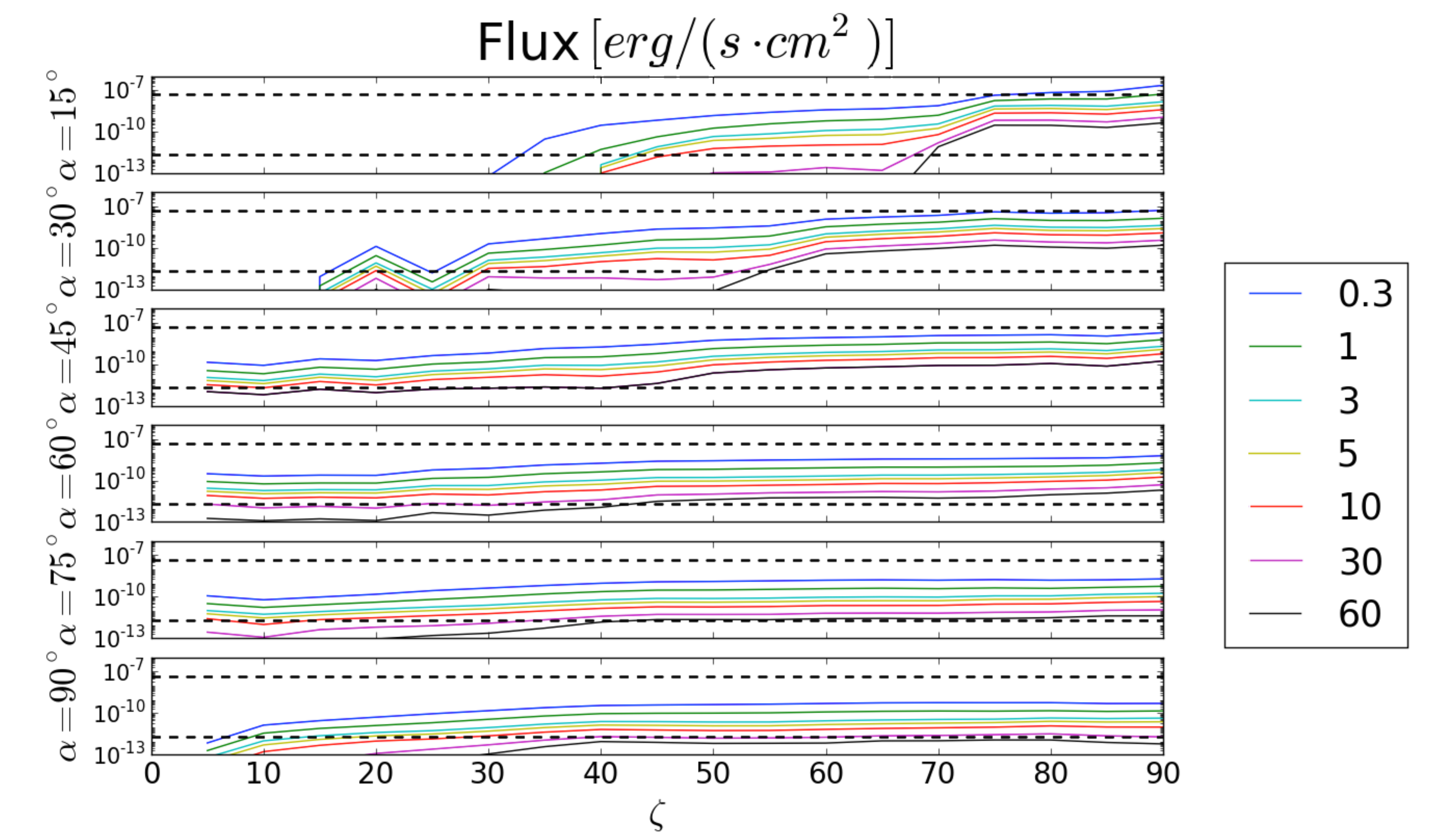}
\caption{The evolution of the fluxes for a pulsar with $P = 0.1s$, $B=2\cdot10^{12}G$ and distance 1kpc for different $(\alpha,\zeta,\sigma)$. 
The six panels are the different $\alpha$ starting from the top to the bottom: $15^\circ, 30^\circ, 45^\circ, 60^\circ, 75^\circ, 90^\circ$.  
Line colors denote different $\sigma$ (in units of $ \Omega$), according to the legend at the side of the plot.  On the
x axis different $\zeta$ from $0^{\circ}$ to $90^{\circ}$; the y axis is the flux in $\rm erg\, s^{-1}
cm^{-2}$. The dashed lines are the minimum and maximum flux measured by \textit{Fermi} for young $\gamma$-ray pulsars, 
rescaled to a distance of 1kpc\label{fluVA}.}
\end{figure*}

We show some figures discussed above illustrating general model properties and the data/model comparison.  For the latter, the first figure for each pulsar shows the observed and model profile and the superimposed phase resolved spectra for $\Gamma$ and $E_{cut}$. The second figure shows the evolution of the observed and model light curve in the energy bins: $100MeV\div 300MeV$, $0.3GeV\div 1$ GeV, $1GeV\div 3$ GeV, $>3$ GeV, $>10$ GeV. On the right are the FIDO predictions, on the left the data.
\section{Vela - J0835-4510}
The Vela pulsar has $P=0.089s$ and $B=3.4\cdot10^{12}G$. The
$\Delta=0.43$, the peaks ``half-height width'' is $0.003$ and
$0.005$. The spectrum has $E_{cut}=3.0$ GeV, $\Gamma=1.5$ and the
flux observed is $9.1\cdot 10^{-9}\,\rm erg/(s\cdot cm^2)$.
\newline
The model candidate has period $P=0.1s$ and $B=4\cdot10^{12}G$. It
is for $\alpha=60^{\circ}$, $\zeta=50^{\circ}$, $\sigma=10 \Omega$. The
$\Delta=0.42$, the peaks ``half-height width'' is $0.06$ and $0.07$
where the accuracy is $\pm0.03$ due to the smoothing. Our candidate
has $E_{cut}=4.1$ GeV, $\Gamma=1.4$ and a flux of $1.9\cdot
10^{-9}\,\rm erg/(s\cdot cm^2)$.

\begin{figure*}
\centering
\includegraphics[width=\textwidth,trim=0 0 0 0]{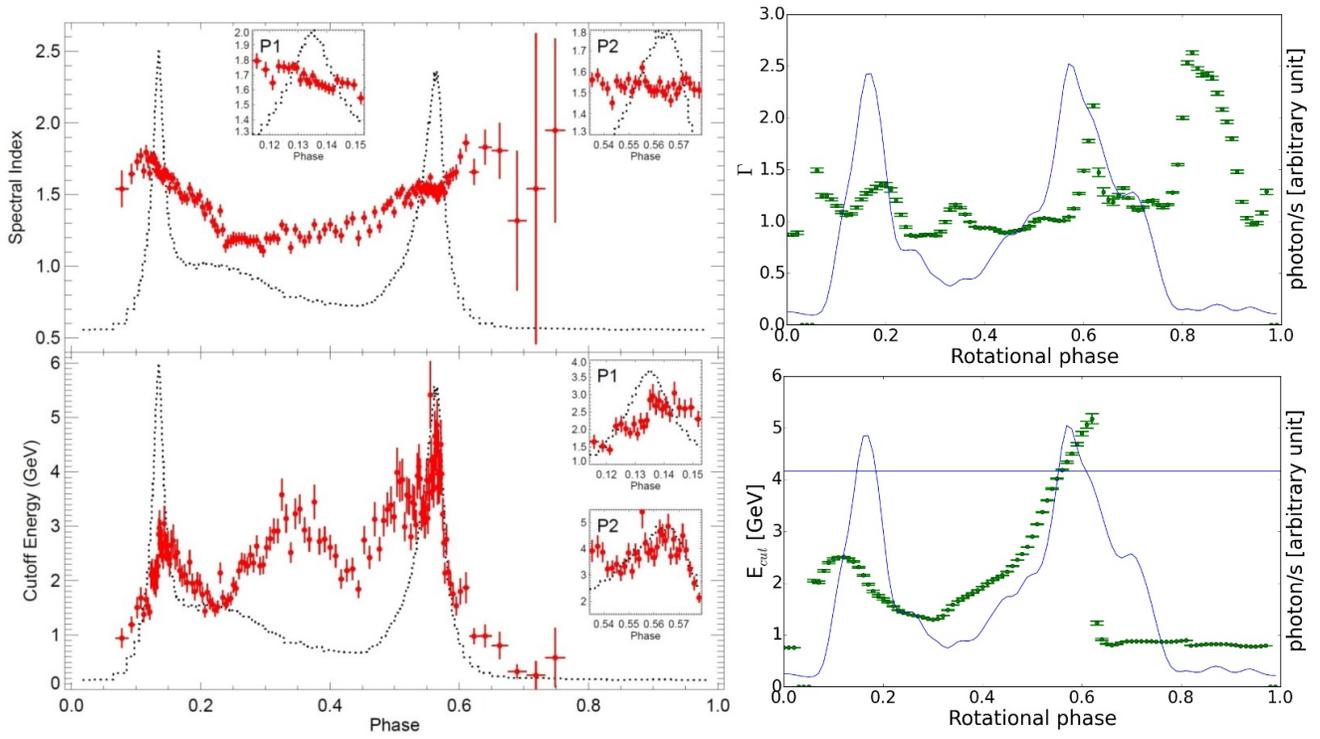}
\caption{Phase resolved spectra for Vela. Left panels: data from
\cite{Megan}. Right panels: model candidate light curves (blue) and
model spectral index (top) and $E_{cut}$ (bottom) (green) as a
function of pulsar phase. The horizontal blue line marks the $E_{cut}$ of
the phase-resolved spectrum.}
\end{figure*}

\begin{figure*}
\centering
\includegraphics[width=\textwidth,trim=0 0 0 0]{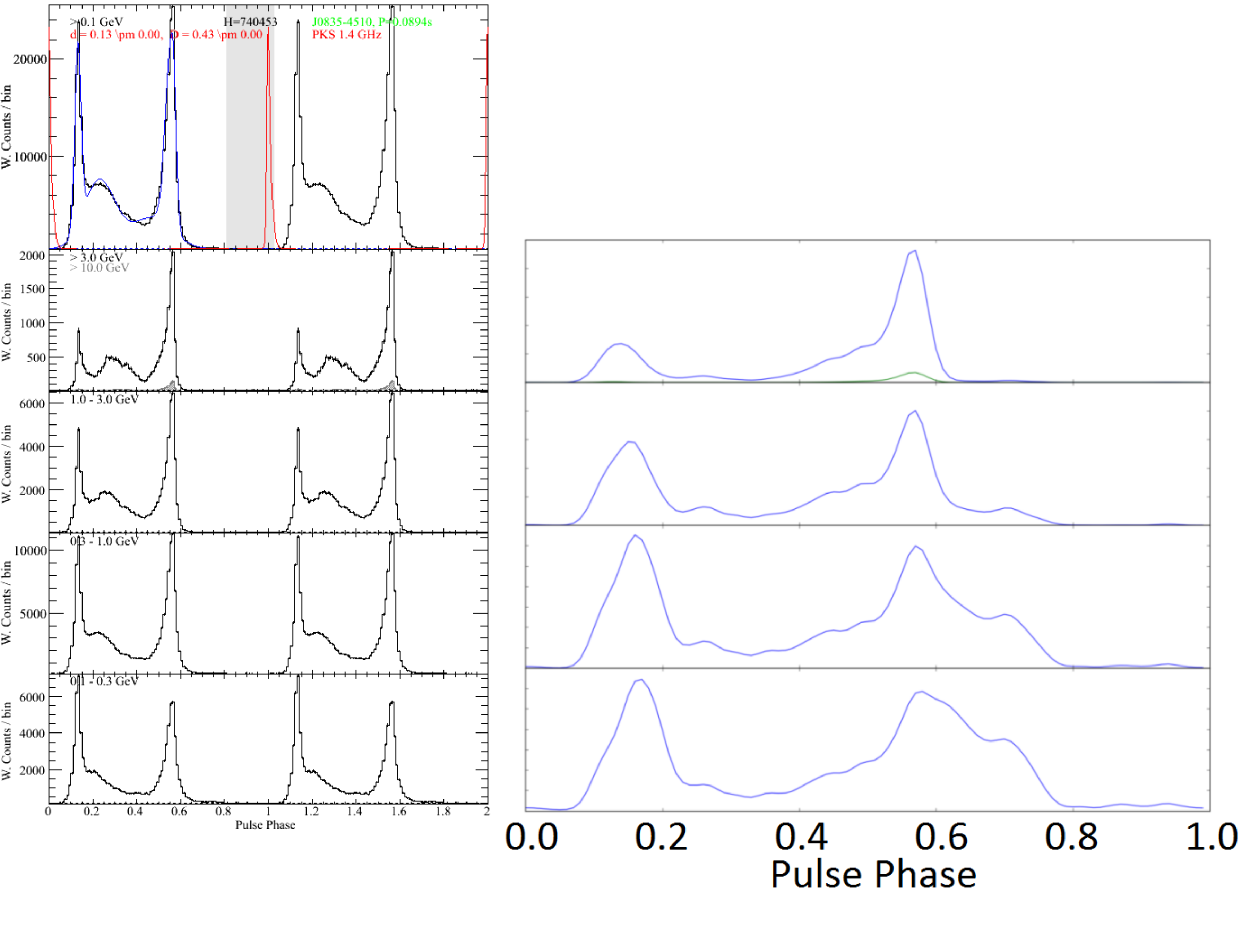}
\caption{Light curve energy evolution for Vela. Left: Observed light
curve in various energy bands, as noted, from \citep{2ndFERMIcat}.
Right: Model light curves for the same energy ranges as the data.}
\end{figure*}

\section{Crab - J0534+2200}
The Crab pulsar has $P=0.033s$ and $B=3.8\cdot10^{12}G$. The
$\Delta=0.40$, the peaks ``half-height width'' is $0.04$ and $0.10$.
The spectrum has $E_{cut}=4.2$ GeV, $\Gamma=1.9$ and the flux
observed is $1.3\cdot 10^{-9}\,\rm erg/(s\cdot cm^2)$.
\newline
The model candidate has period $P=0.03s$ and $B=4\cdot10^{12}G$.
It is for $\alpha=75^{\circ}$, $\zeta=63^{\circ}$, $\sigma=30 \Omega$. The
$\Delta=0.41$, the peaks ``half-height width'' is $0.04$ for the
first peak where the accuracy is $\pm0.03$ due to the smoothing. The
candidate's second peak is too broad or too low. Our candidate has
$E_{cut}=3.4$ GeV, $\Gamma=1.3$ and a flux of $1.6\cdot
10^{-9}\,\rm erg/(s\cdot cm^2)$.

\begin{figure*}
\centering
\includegraphics[width=\textwidth,trim=0 0 0 0]{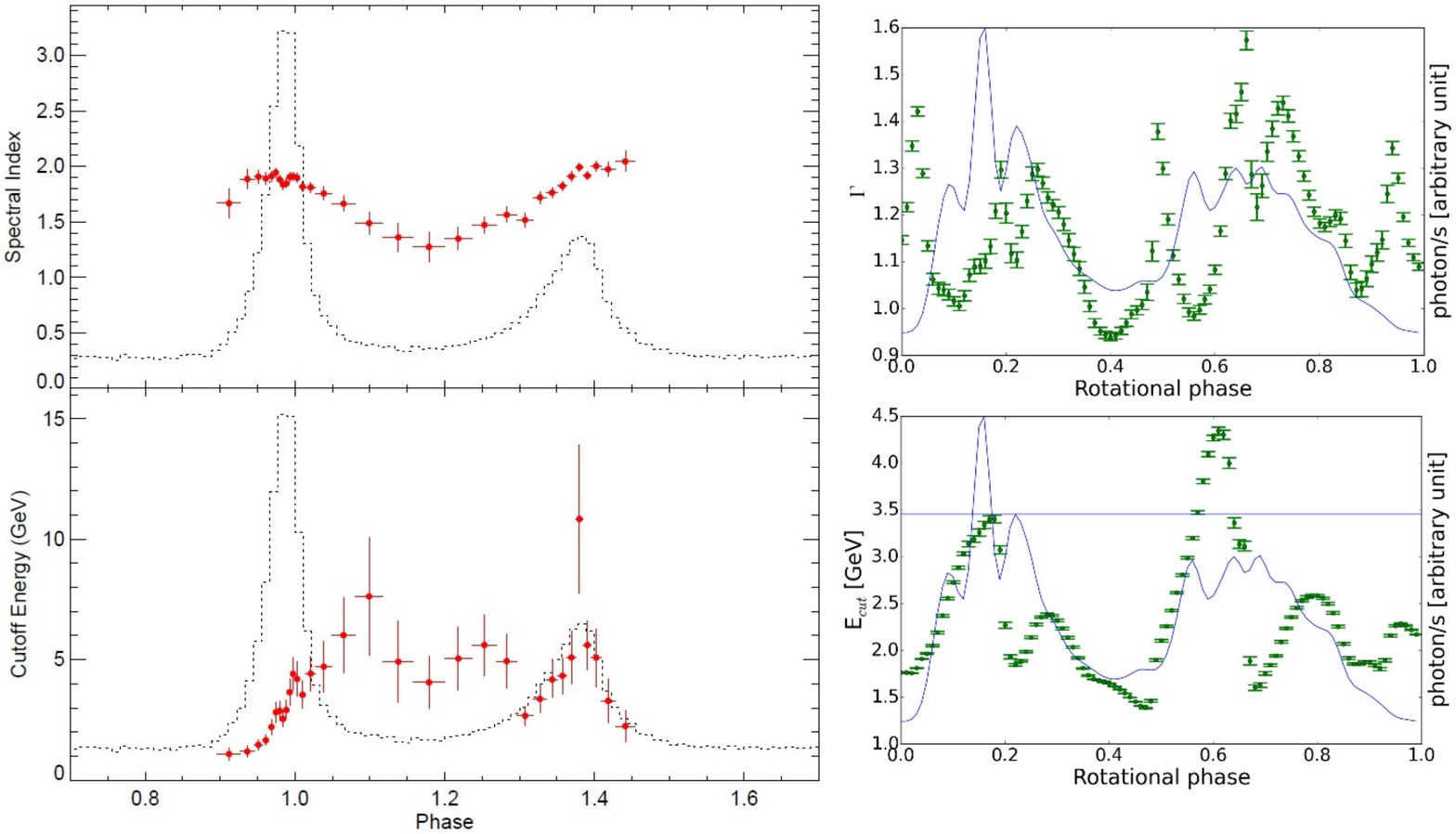}
\caption{Phase resolved spectra for Crab. Left panels: data from
\cite{Megan}. Right panels: model candidate light curves (blue) and
model spectral index (top) and  $E_{cut}$ (bottom) (green) as a function
of pulsar phase.}
\end{figure*}

\begin{figure*}
\centering
\includegraphics[width=\textwidth,trim=0 0 0 0]{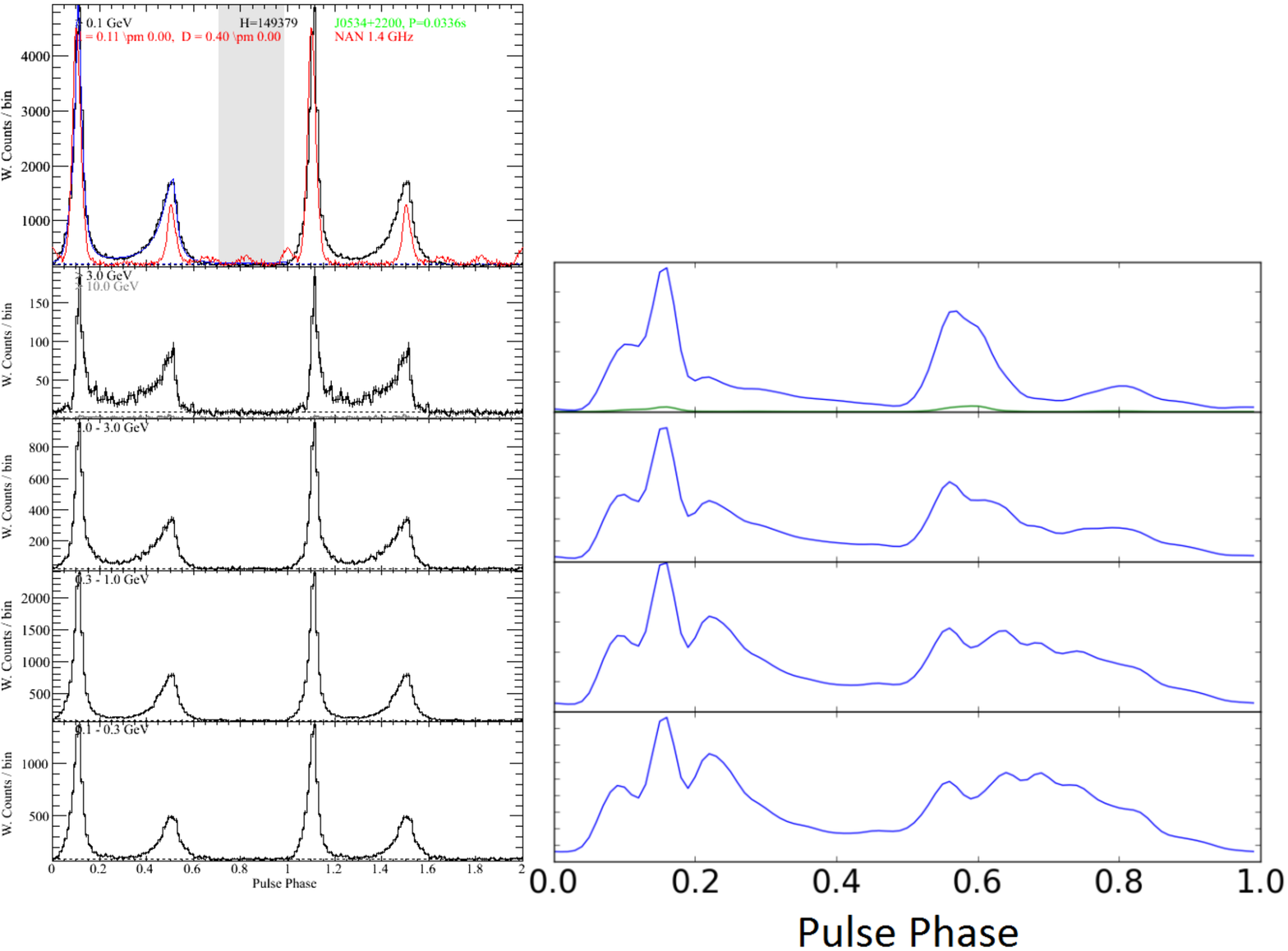}
\caption{Light curve energy evolution for Crab. Left: Observed light
curve in various energy bands, as noted, from \citep{2ndFERMIcat}.
Right: Model light curves for the same energy ranges as the data.}
\end{figure*}

\section{J1952+3252}
J1952 has $P=0.04s$ and $B=0.5\cdot10^{12}G$. The $\Delta=0.48$,
the peaks ``half-height width'' is $0.04$ and $0.1$. The spectrum
has $E_{cut}=2.5$GeV, $\Gamma=1.5$ and the flux observed is
$1.4\cdot 10^{-10}\,\rm erg/(s\cdot cm^2)$.
\newline
The model candidate has period $P=0.03s$ and $B=0.5\cdot10^{12}G$.
It is for $\alpha=75^{\circ}$, $\zeta=85^{\circ}$, $\sigma=10 \Omega$. The
$\Delta=0.5$, the peaks ``half-height width'' is $0.06$ and $0.06$
where the accuracy is $\pm0.03$ due to the smoothing. Our candidate
has $E_{cut}=2.2$ GeV, $\Gamma=1.4$ and a flux of $7.4\cdot
10^{-11}\,\rm erg/(s\cdot cm^2)$.

\begin{figure*}
\centering
\includegraphics[width=\textwidth,trim=0 0 0 0]{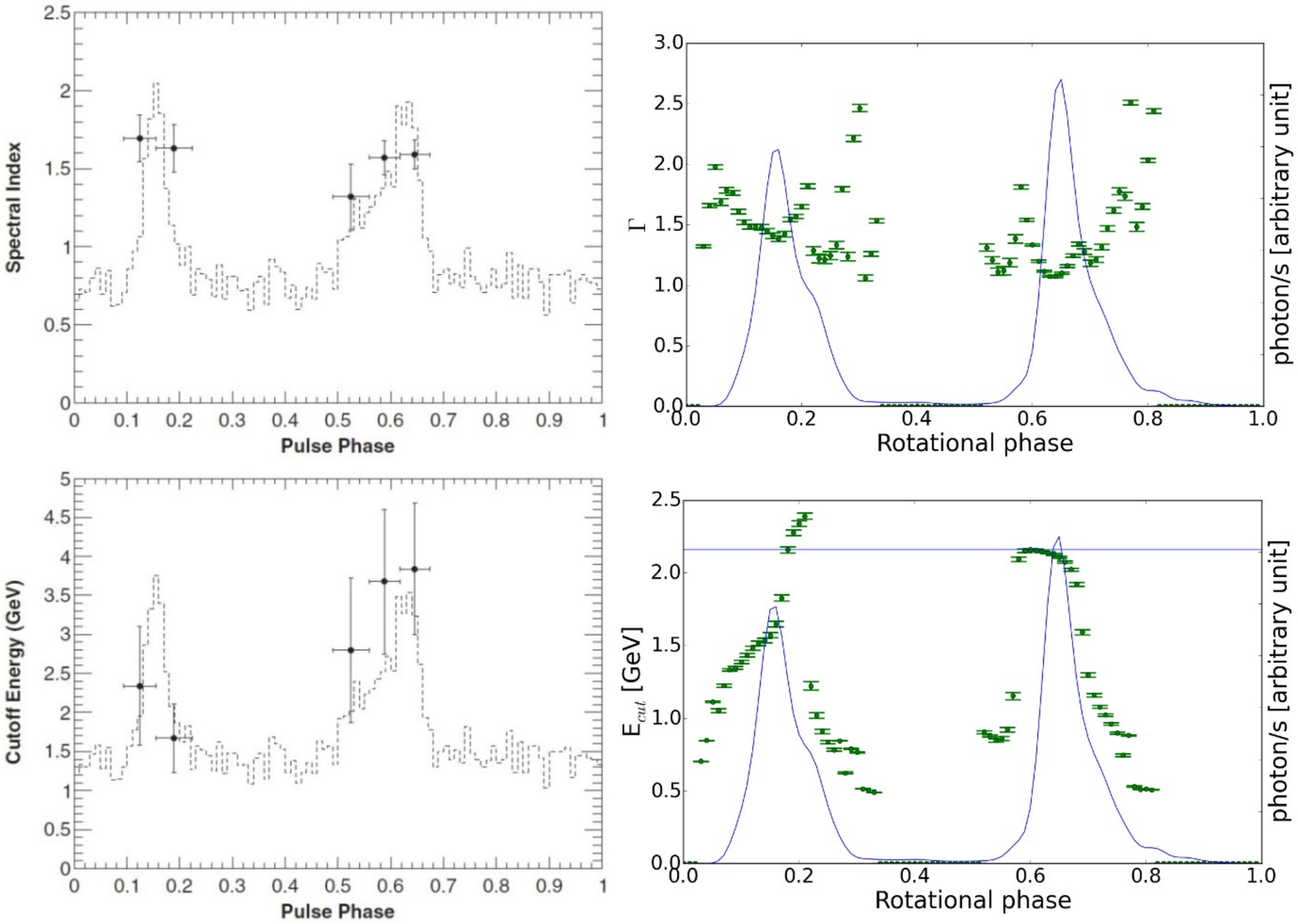}
\caption{Phase resolved spectra for J1952+3252. Left panels: data
from \cite{threePul}. Right panels: model candidate light curves
(blue) and model spectral index (top) and $E_{cut}$ (bottom) (green) as a
function of pulsar phase.}
\end{figure*}

\begin{figure*}
\centering
\includegraphics[width=\textwidth,trim=0 0 0 0]{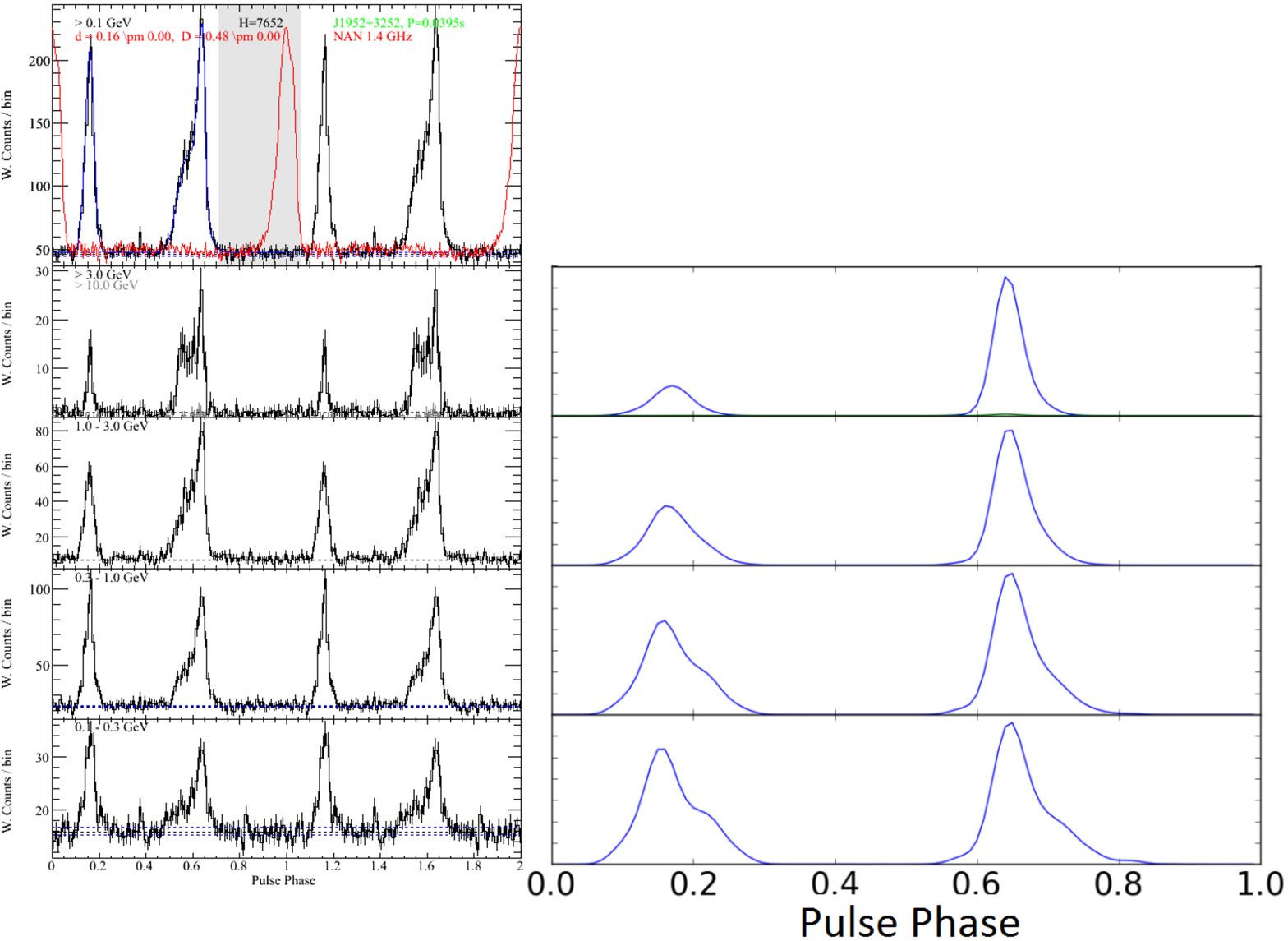}
\caption{Light curve energy evolution for J1952+3252. Left: Observed
light curve in various energy bands, as noted, from
\citep{2ndFERMIcat}. Right: Model light curves for the same energy
ranges as the data.}
\end{figure*}

\end{document}